# Crisis Communication Patterns in Social Media during Hurricane Sandy

Arif Mohaimin Sadri, Ph.D.
Visiting Assistant Professor
Civil and Environmental Engineering
Rose-Hulman Institute of Technology
5500 Wabash Avenue
Terre Haute, IN 47803
Email: sadri.buet@gmail.com
(Corresponding Author)

Samiul Hasan, Ph.D.
Assistant Professor
Department of Civil, Environmental, and Construction Engineering
University of Central Florida
12800 Pegasus Drive, Orlando, FL 32816
Phone: 407-823-2480
Email: samiul.hasan@ucf.edu

Satish V. Ukkusuri, Ph.D.
Professor
Lyles School of Civil Engineering
Purdue University
550 Stadium Mall Drive
West Lafayette, IN 47907, USA
Phone: (765)-494-2296
Fax: (765)-496-7996
Email: sukkusur@purdue.edu

Manuel Cebrian, Ph.D.
Principal Research Scientist
Data61, CSIRO
115 Batman Street, West Melbourne VIC 3003, Australia
Phone: +61 0403 754 676
Email: Manuel.Cebrian@data61.csiro.au

Word Count: 03 tables + 04 figures + 5697 words = 7447 word equivalents

**ABSTRACT**

Hurricane Sandy was one of the deadliest and costliest of hurricanes over the past few decades. Many states experienced significant power outage, however many people used social media to communicate while having limited or no access to traditional information sources. In this study, we explored the evolution of various communication patterns using machine learning techniques and determined user concerns that emerged over the course of Hurricane Sandy. The original data included ~52M tweets coming from ~13M users between October 14, 2012 and November 12, 2012. We run topic model on ~763K tweets from top 4,029 most frequent users who tweeted about Sandy at least 100 times. We identified 250 well-defined communication patterns based on perplexity. Conversations of most frequent and relevant users indicate the evolution of numerous storm-phase (warning, response, and recovery) specific topics. People were also concerned about storm location and time, media coverage, and activities of political leaders and celebrities. We also present each relevant keyword that contributed to one particular pattern of user concerns. Such keywords would be particularly meaningful in targeted information spreading and effective crisis communication in similar major disasters. Each of these words can also be helpful for efficient hash-tagging to reach target audience as needed via social media. The pattern recognition approach of this study can be used in identifying real time user needs in future crises.

**Keywords**: Hurricane Sandy, crisis communication, user concern, topic model, social media, Twitter;

**BACKGROUND AND MOTIVATION**

In 2012, New York and New Jersey coastal residents experienced a massive storm surge produced by late season Hurricane Sandy that caused disastrous affects in the mid-Atlantic and northeastern United States across the Atlantic basin (*1*). Some examples include: $50B in property damage, 72 fatalities, at least 147 direct deaths, destruction of 570K buildings, cancellation of 20K flights, 8.6M power outages in 17 states, 230K cars destroyed due to flooding and thousands of people were displaced from their homes (*2-5*). Disaster resilience is now a national imperative at all levels with a view to limiting such adverse impacts and more efficient policies are required by the government to allow people to be less dependent on federal resources (*6*).

As such, vulnerable communities need to respond to any form of disaster with enough preparation (*7-10*) and effective crisis communication is one major aspect of it. This include systematic planning, collection, organization, and circulation of relevant awareness information to the target audience, reaching out to every individual in a community (*11-13*). During an emergency, people may obtain weather information from traditional media such as radio or television and social media such as Facebook, Twitter, or the internet. Social media platforms, uniquely different from traditional ones, can help disseminate targeted information during a crisis. However, it is critical to efficiently harness the large-scale and rich information from these communication sources (*14*). Many studies have acknowledged such value of social media data in emergency response (*15-20*), community interactions (*21; 22*), crisis informatics (*23-31*), and predicting real world actions/events (*32; 33*).

During Hurricane Sandy, social media also played an important role on crisis communication. New York and New Jersey residents used smart-phones to access social media since they had limited access to traditional information sources (*34*). Social media communications via social media continued during and after the storm in areas without power (*35*). Studies have identified authorities, news media, and peers as emergency warning sources (*36*) and psychological and social factors are very important in translating hazard warning information into a collective decision (*37-41*). For example, local authorities, peers, local and national media, and internet significantly influence evacuation decision-making (*38*). Moreover, individuals were more likely to evacuate if they relied on social media for weather-related information during Sandy (*42*).

Twitter users can post a brief message (maximum 140 characters) and follow other in a network setting. Twitter thus exhibits the characteristics both of a social network and an informational network (*43*). While social network aspects of Twitter provide access to geographically and personally relevant information, the information network can help controlling global information contagion (*33*). These specific features make Twitter particularly useful for effective information dissemination during crises as evidenced in many empirical studies: service characteristics (*17; 26*), retweeting activity (*44; 45*), situational awareness (*46; 47*), online communication of emergency responders (*48; 49*), text classification and event detection (*23; 24; 28; 50; 51*), devise sensor techniques for early awareness (*52*), quantifying human mobility (*53; 54*), and disaster relief efforts (*55*).

In this paper, we analyzed the evolution of communication patterns using machine learning techniques and determined various user concerns that emerged over the course of Hurricane Sandy. The original data included ~52M tweets coming from ~13M users between October 14, 2012 and November 12, 2012. We run topic model on ~763K tweets from top 4,029 most frequent users who tweeted about Sandy at least 100 times. We identified 250 well-defined communication patterns based on perplexity. Conversations of most frequent and relevant users indicate the evolution of numerous storm-phase (warning, response, and recovery) specific topics. People were also concerned about storm location and time, media coverage, and activities of political leaders and celebrities. We also present each relevant keyword that contributed to one particular pattern of user concern.

**DATA DESCRIPTION AND ANALYSES**

In this study, we analyze raw data (~52 M tweets, ~13 M users, Oct 14 -Nov 12, 2012) obtained from Twitter. Please see (*52*) for the detailed steps involved in data collection. The data includes a text database with user and text identifiers, texts, and some additional useful information. The network database includes the relationship graphs of active users i.e. the list of followees for each user (*31*). These were reconstructed using Twitter API. Only a minor fraction of the texts (~ 1.35%) are geo-tagged by Twitter. For relevance, user activity was assessed on the basis of number of tweets (~11.83 M) in the data that included the word 'sandy', co-appeared with other words after filtering out ~46.45 M tweets that are in English i.e. non-English tweets were removed.

First, we analyzed the texts that evolved in the tweets of top 4,029 most frequent users who tweeted at least 100 times about Sandy (AF ≥ 100), producing ~763 K tweets and ~95 M words. After removing the punctuations, English stop-words, repeated words such as 'http' and '@' from these tweets, we observed that the 50 most frequent words contribute ~30% of all the words in the texts. We present the highly frequent words as a word cloud in Figure 1 where the frequencies of these words can be seen. 'sandy' and 'hurricane' are the top two most frequent word. The word 'storm' is also on top of the list. Interestingly, we found that the word 'storm' was more popular on October 23, 2012 shortly after the tropical depression was formed (Table 1). Later on, 'hurricane' and 'sandy' became more frequent close to the landfall on October 29. The word 'superstorm' was also popular. The appearance of 'sandy' before the tropical depression was formed (October 22) suggests those tweets having 'sandy' out of context. The labelling of a tropical storm should be given importance as it is strongly connected to how information disseminates later. Any given word can be a part of a specific topic evolved over time. Also, each word can be a part of multiple topics. The text database needs to be carefully analyzed to infer aggregate and user-specific topic pattern.

We also analyzed the texts that evolved in the tweets of the top 157,622 most frequent users who tweeted at least 10 times about Sandy (AF ≥ 10), producing ~3.9 M tweets. After removing the punctuations, English stop-words, repeated words such as 'http' and '@' from these tweets, we observed that these 100 most frequent words contribute ~37.5% of all the words in the texts. Any given word can be a part of a specific topic evolved over time. Also, each word can be a part of multiple topics. The text database needs to be carefully analyzed to infer aggregate and user-specific topic pattern. We present the highly frequent words as a heatmap over time in Figure 2 and Figure 3 where the frequencies of these words can be seen. 'sandy' and 'hurricane' are the top two most frequent word. The word 'storm' is also on top of the list. Interestingly, we found that the word 'storm' was more popular on October 23, 2012 shortly after the tropical depression was formed. Later on, 'hurricane' and 'sandy' became more frequent close to the landfall on October 29. The word 'superstorm' was also popular. The appearance of 'sandy' before the tropical depression was formed (October 22) suggests those tweets having 'sandy' out of context. The labelling of a tropical storm should be given importance as it is strongly connected to how information disseminates later. The other words in the list can be broadly classifies into several categories. Words such as 'east', 'coast', 'stay', 'safe', 'update', 'watch', 'path', 'prepare', 'emergency' appeared more frequently before the storm. Some words, for example, 'power', 'water', 'food', 'gas', 'home' were more prominent close to or shortly after the landfall. Considerably after the landfall, topics including 'help', 'relief', 'victims', 'donate', 'affect', 'please', 'fema', 'without', 'shelter' appeared to emerge. Words such as 'weather', 'state', 'time', 'still', 'sandys', 'hit', 'ny', 'nj', 'york' were fairly uniformly distributed both before and after the landfall.

## MODEL DESCRIPTION AND FORMULATION

The crisis communication patterns and their evolution in these networks were based on topic models that are well-established in the machine learning literature. A topic model is a generative model that follows a probabilistic process for generating documents based on a set of straightforward probabilistic sampling rules. This process explains how words in a document can be generated based on some latent variables i.e. topics that evolve in a document. The overall procedure includes the following steps (*56*): (i) select a distribution over topics to make a new document, (ii) for each word in the document, select a topic randomly and then a word from that topic following that distribution. The process can also be reversed and the set of topics that generate the collection of documents can be obtained. The model fitness of such generative models should find the best set of latent variables (i.e. topics of the documents) based on which the observed data can be reasonably explained (i.e. words in the documents). The generative process, as discussed above, does not assume any specific ordering of the words ('bag-of-words' assumption in natural language processing) in a document (*56*) and the frequency of word appearance in a document is the only information that is relevant. The ordering of the words can be useful at times, however, this is not captured by topic models. A topic model can also be applied to other types of discrete data.

Formally speaking, the problem of identifying various social interaction topics is to determine $K$ latent patterns through $\phi_k$ for $k \in \{1,2,...K\}$ and each topic $\phi_k$ is a distribution of different words. A word is defined as the basic unit of data to be selected from a set of possible words of size $W$, a user tweets about $N$ topics, and the user can contribute to the collection of $M$ tweets. The generative process is summarized below (see Figure 4a):

1. For each topic, $k \in \{1,2,...K\}$, a distribution over words is selected
$$\phi^{(k)} \sim \text{Dirichlet}(\beta)$$
2. For each tweet, $m \in \{1,2,...M\}$,
   a) A distribution over topics is selected
   $$\theta^{(m)} \sim \text{Dirichlet}(\alpha)$$
   b) For each word $i$ in tweet $m$
      i. Select a topic $z_i \sim \text{Multinomial}(\theta^{(m)}); z_i \in \{1,2,...K\}$
      ii. From topic $z_i$, a word is selected
      $$w_i \sim \text{Multinomial}(\phi^{(z_i)}); w_i \in \{1,2,...W\}$$

Now, given $M$ tweets, $K$ topics over $W$ unique words, the main objectives of the inference of topic pattern classifications are:

i. Find the probability of a word $w$ given each topic $k$, $P(w|z = k) = \phi_k^w$ where $P(w|z = k)$ is represented with $K$ multinomial distributions $\phi$ over words of size $W$

ii. Find the probability of a topic $k$ for a word in tweet $m$, $P(z = k|m) = \theta_m^k$ where $P(z|m)$ is represented with $M$ multinomial distributions $\theta$ over $K$ topics.

The above model views the topic pattern as a probability distribution over words and tweeting activities as a mixture of these patterns. From $K$ topics, the probability of **i**-th word in a given tweet $m$ is:

$$P(w_i|m) = \sum_{j=1}^{K} P(w_i|z_i = j)\, P(z_i = j|m) \tag{1}$$

Here, $z_i$ is the latent variable referring to the pattern from which the **i**-th word is drawn, $P(w_i|z_i = j)$ indicates the probability of word $w_i$ under the **j**-th pattern and $P(z_i = j|m)$ is the probability of selecting a word from pattern **j** in the tweet $m$. Intuitively, $P(w|z)$ determines the importance of a word in forming a pattern and $P(z|m)$ determines the prevalence of the pattern in different tweets. The complete model of pattern generation by IPM follows:

$$w_i|z_i, \phi^{(z_i)} \sim \text{Multinomial}\left(\phi^{(z_i)}\right)$$
$$\phi \sim \text{Dirichlet}(\beta)$$
$$z_i|\theta^{(m)} \sim \text{Multinomial}\left(\theta^{(m)}\right)$$
$$\theta \sim \text{Dirichlet}(\alpha)$$

Here, $\alpha$ and $\beta$ are hyper-parameters for the prior distributions of $\theta$ and $\phi$ respectively. We assume Dirichlet prior distributions which are conjugate to the multinomial distributions.

The joint distribution of words and patterns $P(\mathbf{w}, \mathbf{z})$ written as:

$$P(\mathbf{w}, \mathbf{z}) = P(\mathbf{w}|\mathbf{z})\, P(\mathbf{z}) \tag{2}$$

The first term can be written as (*57*):

$$P(\mathbf{w}|\mathbf{z}) = \left(\frac{\Gamma(W\beta)}{\Gamma(\beta)^W}\right)^K \prod_{k=1}^{K} \frac{\prod_w \Gamma(n_k^w + \beta)}{\Gamma(n_k^{(\cdot)} + W\beta)} \tag{3}$$

Here $n_k^w$ is the number of times word $w$ is assigned to pattern $k$ and $n_k^{(\cdot)} = \sum_{w=1}^{W} n_k^w$. The second term can be written as:

$$P(\mathbf{z}) = \left(\frac{\Gamma(K\alpha)}{\Gamma(\alpha)^K}\right)^M \prod_{m=1}^{M} \frac{\prod_k \Gamma(n_m^k + \alpha)}{\Gamma(n_m^{(\cdot)} + K\alpha)} \tag{4}$$

Here $n_m^k$ is the number of times a word from tweet $m$ is assigned to pattern $k$ and $n_m^{(\cdot)} = \sum_{k=1}^{K} n_m^k$. A pattern can be assigned to a word using the following conditional distribution (*57*) :

$$P(z_i = k|\mathbf{z}_{-i}, \mathbf{w}) \propto \frac{n_{-i,k}^{w_i} + \beta}{n_{-i,k}^{(\cdot)} + W\beta} \cdot \frac{n_{-i,m_i}^k + \alpha}{n_{-i,m_i}^{(\cdot)} + K\alpha} \tag{5}$$

Here, $n_{-i}$ is the count excluding the current pattern assignment of $z_i$. The first ratio expresses the probability of word $w_i$ in pattern $k$ and the second ratio expresses the probability of pattern $k$ in the tweet $m$.

## PARAMETER ESTIMATION AND MODEL SELECTION

There are various approximation techniques for estimating the parameters of this model (*57; 58*). We used the Gibbs sampling approach proposed by (*57*). The algorithm can be found in detail in (*57*). Only a brief description of the approach is provided here. To estimate the model parameters, a Markov Chain Monte Carlo (MCMC) procedure is used. In MCMC, samples are taken from a Markov chain constructed to converge to a target distribution. In our model, each state of the chain is the assignment of a pattern to a word and the transition from one state to another follows a specific rule based on Gibbs sampling approach (*59*). In this procedure, the next state is reached by sampling the variables from a conditional distribution which specifies the distribution of the variables conditioned on the current assignment of all other variables and the observations. The parameters of LDA, representing the hidden patterns, can be computed as:

$$\hat{\phi}_k^w = \frac{n_k^w + \beta}{n_k^{(\cdot)} + W\beta} \; ; \; \hat{\theta}_m^k = \frac{n_m^k + \alpha}{n_m^{(\cdot)} + K\alpha} \tag{6}$$

The model selection was based on *Perplexity* – a metric to measure the predictive capacity of the model to infer the unseen data in each run. In machine learning, *Perplexity* is a commonly used metric to report the performance of a probabilistic model that refers to the average likelihood of obtaining a test data set given a set of model parameters. *Perplexity* can be defined as the exponential of the negative of average predictive likelihood of a test data given a model (*57*). In this study, the algorithm was run for different number of topic patterns (K) and perplexity in each run was computed. Next, the optimal number of patterns was selected based on perplexity values. For a given set of words $\{w_m\}$ and $m \in D^{test}$ given a model $\mathcal{M}$, *Perplexity* of a test data set can be defined as:

$$Perplexity = \exp\left[-\frac{\sum_{m=1}^{M} \log p(w_m | \mathcal{M})}{\sum N_m}\right] \tag{7}$$

where $N_m$ is the number of words in each topic $m$ and $p(w_m | \mathcal{M})$ can be derived from Eq. (1). Figure 4.b indicates that there exist 250 well-defined pattern in the data since there is no significant change in perplexity after that.

**FIGURE 1: Patterns obtained from the Text Analysis** (AF ≥ 100; 4,029 nodes/users and 763,000 tweets)

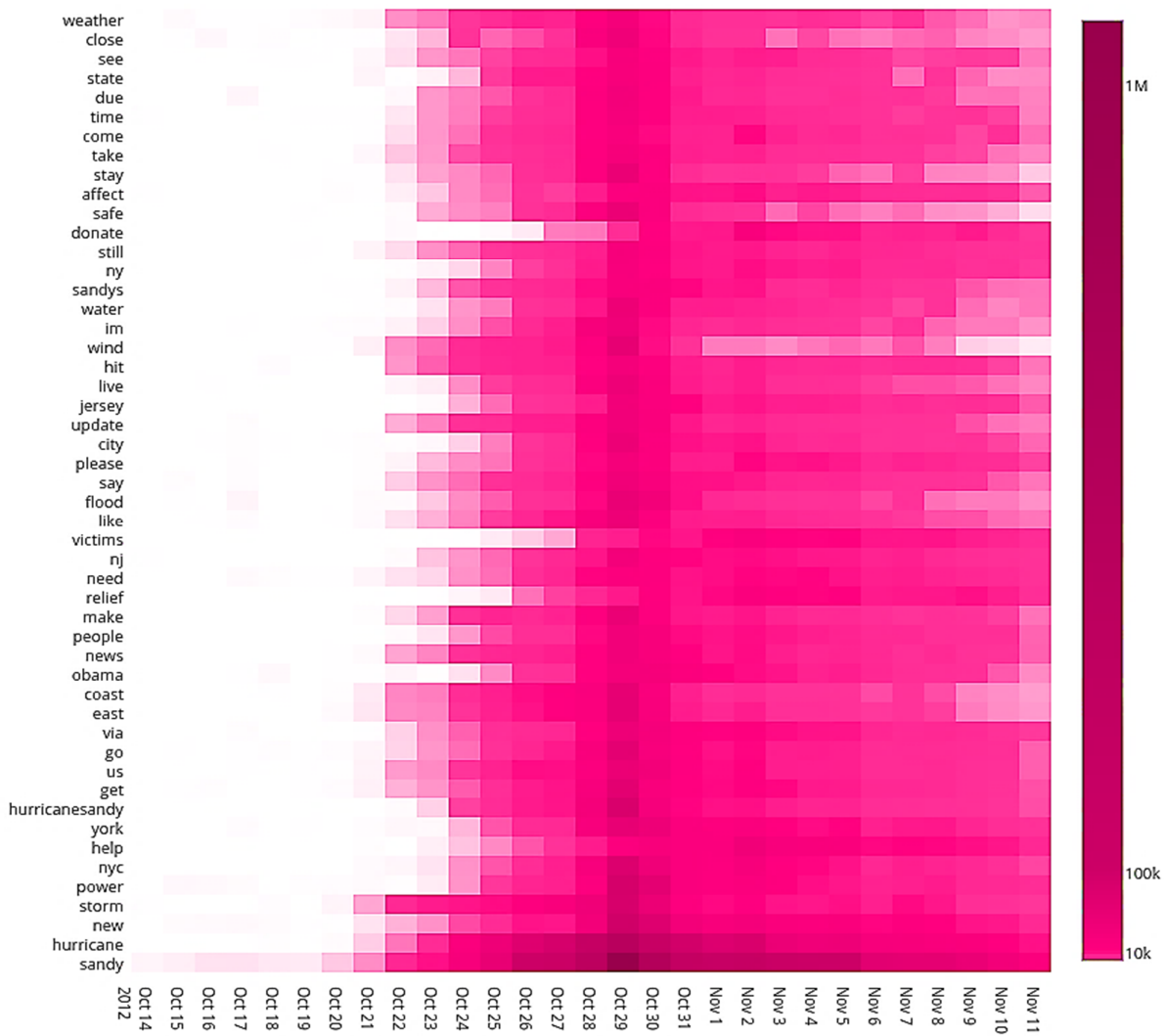

**FIGURE 2 Word appearance over time (word frequency rank: 1-50)**

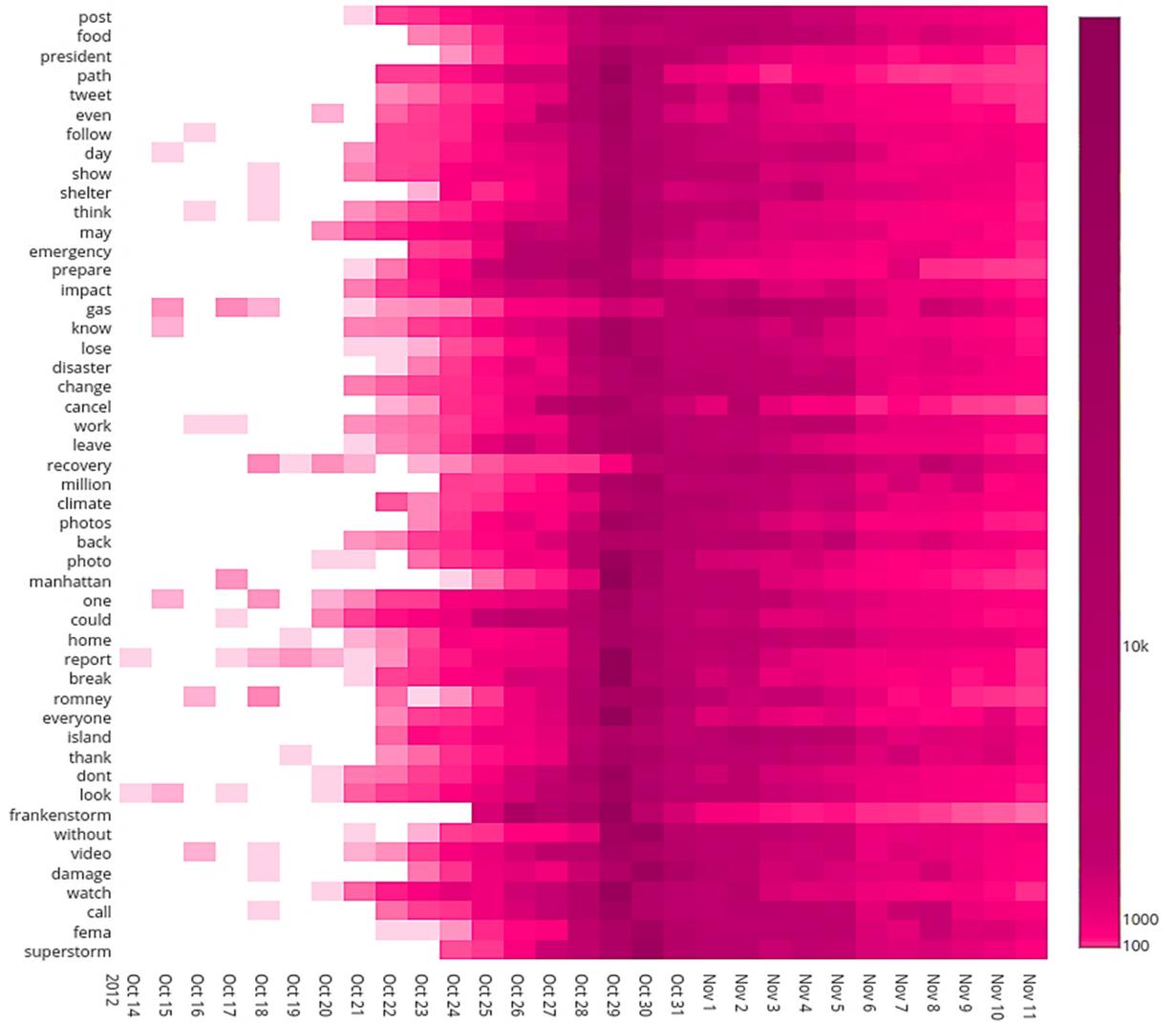

**FIGURE 3 Word appearance over time (word frequency rank: 51-100)**

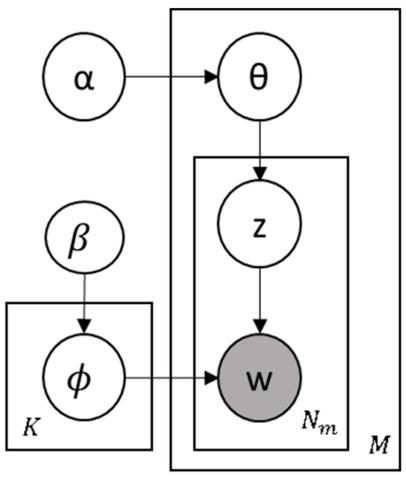

(a)

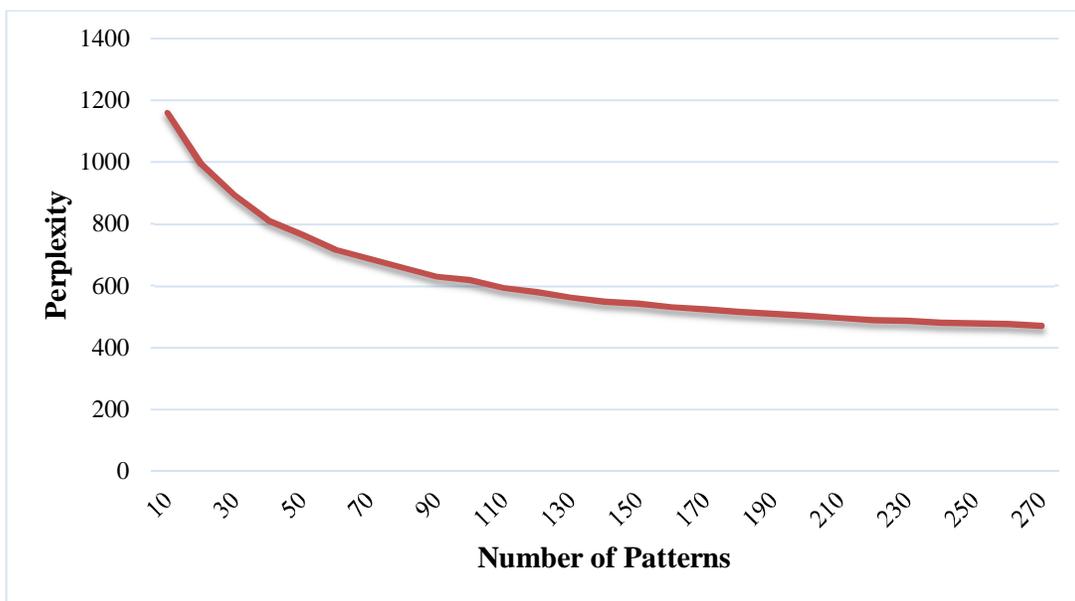

(b)

**FIGURE 4 (a) Probabilistic generative process of topic models (b) Perplexity Analysis**

**TABLE 1 Communication Patterns: Warning and Recovery Phase**

| | *Topic* | *No. of Topics* | *Pattern #* | *Relevant Keywords* |
|---|---|---|---|---|
| **Warning Phase** | Storm prediction | 13 | W1-13 | model, track, show, predict, impact, data, confirm, intense, image, satellite, latest, radar, nasa, loop, powerful, view, cuba, jamaica, hurricane, might, become, frankenstorm, increase, storm, surge, lifethreatening, flood, graphics, make, landfall, expect, along, threaten, become, merge, tropical, strength, downgrade, form, cyclone, gain, weaken, regain, region, hit, path, across, map, crisis, gouge, noaa, see, direct |
| | Storm watch | 6 | W14-19 | warn, storm, watch, effect, issue, tropical, huge, hit, eye, keep, unclear, video, space, show, nasa, station, timelapse, natioanl, hurricane, center, forecast, nhc, advisory, cone |
| | Preparedness | 4 | W34-37 | tip, disaster, children, care, preparedness, parent, kid, prepare, residents, urge, plan, officials, get, ready, try, serious, make, sure, check, neighbor, elderly |
| | User concern | 8 | W38-45 | affect, thoughts, prayers, everyone, pray, god, love, hope, feel, worry, call, shit, take, seriously, aim, place, care, action, warn, stay, safe, keep, please, inside, tune, inform, impact, wake, destructive, know, need, want, anyone, let, call, someone, put, risk, really, info, visit, check, control, rumor, share |
| | Weather condition | 9 | W20-28 | wind, rain, high, heavy, strong, snow, mph, move, category, sustain, mb, pressure, water, level, tide, feet, rise, surge, record, dangerous, outside, inside, scary, gust, far, miles, force, storm, center, field, wide, air, drop, eye, low, central, inch, report, blizzard, part |
| | Previous hurricane | 2 | W29-30 | review, katrina, unprepared, usgov, irene, tragedy, vulnerability |
| | Storm cause | 3 | W31-33 | blame, gay, marriage, preacher, global, warming, climate, change, cause |
| **Recovery Phase** | Disaster Relief | 16 | RC1-16 | concert, victims, raise, relief, benefit, sandyaid, occupysandy, sandyvolunteer, sandyrelief, sandyhelp, mutualaid, donate, redcross, help, efforts, fund, support, million, show, need, church, volunteer, help, pls, collect, please, recover, rt, spread, word, families, charity, ask, offer, free, aid, provide, displace, want, supply, join, dayofgiving, fundraise, disaster, federal, fema, blood, drive, fdr, food, meals, emergency, serve, response |
| | Recovery | 10 | RC17-26 | reopen, jfk, airport, help, survivors, victims, impact, cash, recovery, efforts, begin, fuel, frustration, please, need, help, family, claim, pay, insurance, assistance, fema, tax, health, workers, mental, cut, learn, lessons, get, back, work, normal, finally, replacement, recover, add, businesses |
| | Damage/Aftermath | 7 | RC27-33 | damage, cost, billion, estimate, economic, losses, aftermath, wake, wrath, fury, devastation, devstate, fee, waive, bank, ever, history, record, largest, worst, biggest, sales, auto, industry, affect, buy, please, help, areas, concern, postsandy, expose |

**TABLE 2 Communication Patterns: Response Phase**

| | *Topic* | *No. of Topics* | *Pattern #* | *Relevant Keywords* |
|---|---|---|---|---|
| **Response Phase** | Gas/fuel | 3 | RP1-3 | fuel, spill, diesel, oil, gasoline, near, shut, gas, line, station, ration, price, shortage, wait |
| | Food/water | 2 | RP4-5 | food, water, need, supply, blanket, batteries, distribute, clothe, ice, drink, pump, bottle, boil, safe, sewage |
| | Power outage | 11 | RP6-16 | power, without, million, customers, still, remain, restore, utility, phone, charge, cell, internet, mobile, battery, explosion, dark, safety, generator, caution, blackout, tip, cold, house, electricity, outage, without |
| | Transportation | 9 | RP17-25 | flight, travel, tour, airlines, subway, transit, mta, commuter, bike, walk, bus, service, train, cancel, suspend, system, line, resume, suspend, traffic light, break, stop, wait, port, metro, authority, bridge, tunnel, open, close, station, boat, empty, flood |
| | Local officials | 3 | RP26-28 | officials, governor, cuomo, christie, obama |
| | Evacuation | 1 | RP29 | evacuation, zone, order, mandatory |
| | Infrastructure | 3 | RP30-32 | statue, liberty, nuclear, power, plant, world, trade, center |
| | Death tolls | 5 | RP33-37 | dead, death, rise, toll, destruction, homeless, kill, people, many, lot, bahamas, haiti, caribbean, fear, video |
| | Hospital | 1 | RP38 | hospital, evacuate, medical, patient, nurse, fail, baby |
| | Rescue | 2 | RP39-40 | police, people, nypd, rescue, fire, trap, loot, report, ship, miss, sink |
| | Fire | 1 | RP41 | home, fire, destroy, burn |
| | Flood | 2 | RP42-43 | flood, street, swim, severe, underwater |
| | Trees/Debris | 2 | RP44-45 | cleanup, clean, clear, debris, house, tree, fall, home, block, line, wire |
| | Pets/Animals | 2 | RP46-47 | pet, shelter, rescue, animal, find, help, dog, cat |
| | Crime | 1 | RP48 | worry, crime, strand, unplug |
| | Stock Market | 1 | RP49 | stock, market, trade, close, exchange, open, close |
| | FEMA | 3 | RP50-52 | fema, response, respond, call, 911, register, insurance, damage, shelter, zip, code |
| | First Responder | 6 | RP53-58 | national, guard, navy, aid, response, assist, soldier, army, support, carrier, american, red, cross, donation, million, relief, first, responder, team, supply, military, truck, community, aid, group |
| | Event Cancellation | 4 | RP59-62 | cancel, due, event, class, marathon, runner, suspend, office, postpone, date |
| | Work/ School Colsure | 2 | RP63-64 | school, close, reopen, remain, due, office |

**TABLE 3 Communication Patterns: Location, Time, News & Media, Politics, Leaders & Celebrities, and Others**

| | *Topic* | *No. of Topics* | *Pattern #* | *Relevant Keywords* |
|---|---|---|---|---|
| **Location** | East Coast | 3 | L1-3 | easterm, east, coast, us |
| | NY/NJ | 4 | L4-7 | ny, nj, new, york, city, jersey, state |
| | NY | 3 | L8-10 | new, york, yorkers, newyork, nyc, nysandy |
| | NJ | 2 | L11-12 | nj, new, jersey, sandynj, njsandy |
| | Other States | 9 | L13-21 | li, long, island, longisland, staten, statenisland, borough, rhode, si, ct, connecticut, ctsandy, va, virginia, vasandy, nc, north, carolina, delaware, philadelphia, philly, sandycenpa, md, maryland, mdsandy |
| | River | 1 | L22 | river, hudson, bank |
| | Street Address | 1 | L23 | st, ave, cars, water, flood, avenue, street |
| **Politics, Leaders & Celebrities** | President Obama | 3 | PC1-3 | obama, gop, govt, america, president, barack, statement, speak, brief, deliver |
| | Others | 4 | PC4-7 | romney, mitt, bill, clinton, government, capital, mayor, endorse |
| | US Election | 5 | PC8-12 | presidential, election, us, race, politics, gop, washington, vote, poll, voters, early, election2012 |
| | Celebrities | 4 | PC13-16 | karl, rove, lindsay, lohan, dina, lohan, jim, cantore |
| **News & Media** | News Media | 16 | NM1-16 | cnn, los, angeles, times, sky, news, nyt, business, abc, daily, mail, cbs, usa, today, box, break, post, blog, report, washington, huffington, yahoo, cover, buzzfeed, bbc, wsj, wall, street, journal, weather, channel, national, coverage, forbes, reuters |
| | Social Media | 9 | NM17-25 | social, media, internet, share, friends, photo, video, app, download, mobile, tweet, send, message, text, receive, facebok, fb, page, status, topic, top, twitter, follow, info, information, picture, image, pics, instagram |
| | Broadcasting | 5 | NM26-30 | live, update, coverage, watch, brief, press, conference, hold, stream, tv, online, broadcast, youtube, chat, report, call, ask, question, answer, reporters |
| **Time** | Time | 6 | T1-6 | oct, 29, 30, 2012, pm, tonight, afternoon, night, week, year, last, next, hours, days, minutes, past, next |
| **Others** | Others | 33 | O1-33 | hurricane, sandy, dont, think, cant, u, cant, look, one, superstorm, say, good, ppl, like, get, give, hear, us, go, yet |

## CRISIS COMMUNICATION PATTERNS

### Patterns of Warning Phase

Table 1-3 present the details on each type of the communication pattern along with the associated keywords. The top ten words contributing to each pattern are tabulated in Appendix (Table 1- Table 8) in the decreasing order of their occurrence probability from left to right (*60*). Topic model was applied to ~763 K tweets including ~95 M words from the 4,029 users being active at least 100 times (AF $\geq$ 100) demonstrating their high degree of appearance and relevance to Sandy. Most discussions during the warning phase (before October 29, 2012), listed in Table 1 are related to storm prediction (W1-13), storm watch (W14-19), and storm preparedness (W34-37). Storm prediction topics consist of words like NASA, NOAA, satellite, predict, path, model, track, form, regain, weaken and other similar words. Conversations on storm watch are expressed through keywords including warn, watch, issue, advisory, forecast. People were also concerned about storm preparedness and words like 'get', 'ready', 'try', 'prepare', 'parent', 'children', 'care', 'neighbor' contributed to such topics. A number of topics are specific to user concern (W38-45), people expressed their thoughts and prayers for the people who stayed along the path of the hurricane. Some topics were related to updates on weather condition made of 'wind', rain', 'snow', 'water', 'heavy', 'strong' and several weather measurement units. People also expressed their concern about previous hurricanes such as Katrina and Irene as Sandy was approaching. The causality of a major hurricane like Sandy was also a major topic of concern (W31-33). Some claimed global warming and climate change to be responsible, some supposed gay marriage to be a source of storm formation.

### Patterns of Response Phase

Turning to topics evolved during the response phase (during or immediately after the landfall on October 29, 2012), we observe that people were mainly concerned about the basic needs such as gas/fuel (RP1-3), food/water (RP4-5), and the adverse impacts caused by significant power outage in different states (RP6-16). Several discussions went on transportation related topics (RP17-25) as expressed by 'flight', 'travel', 'airlines', 'subway', 'bus', 'train', 'metro', 'station', 'bridge', 'tunnel', 'mta', 'cancel', 'service', 'suspend', 'resume', 'line' and other similar words (Table 2). People also talked about how local officials such as state governors responded (RP26-28) to the situation and whether or not an evacuation order is issued (RP 29). Major infrastructures such as Statue of Liberty, World Trade Center and Nuclear Power Plant were also part of many interactions (RP30-32). The severity of death tolls was highlighted in several topics (RP33-37). Hospitals undertook precarious evacuations (RP38) and several rescue efforts went on by the police department (RP39-40). Concerns about the impact of fire (RP41), flood (RP42-43), cleaning debris (RP44-45), and pets/animals (RP46-47) also emerged. People also talked about different ongoing crimes (RP48) and the status of the stock market (RP49). First responders such as FEMA (RP50-52), National Guard, Navy, American Red Cross and Army also appeared in a number of topics (RP53-58). Some people discussed about the cancellation of major events (RP59-62) and closure of offices and schools (RP63-64).

### Recovery Phase and Phase Independent Patterns

The recovery phase of the storm primarily included discussions about numerous disaster relief efforts such as Occupy Sandy and popular hashtags like 'sandyaid', 'sandyrelief',

'sandyhelp', 'sandyvolunteer' among others (RC1-16). Some discussions were specific to the recovery process itself when people talked about survivors and victims of the storm and lessons learned from Hurricane Sandy (RC17-26). Economic losses, devastating damage, and other aftermaths caused by Sandy (RC17-26) were also discussed in this particular phase of the storm (Table 1). Various topics were phase independent and specific to location, time, media coverage, political leaders and celebrities as listed in Table 3. When the location category is considered, we observe that many patterns represent user concern about East Coast, New York and New Jersey (L1-12). These were the areas that experienced major impact of Sandy. People also communicated about the other states/areas slightly impacted or close to the path of Sandy. These included Long Island, Staten Island, Rhode Island, Connecticut, Virginia, North Carolina, Delaware, Philadelphia, and Maryland (L13-21). Other specific locations such as Hudson River (L22) and any given street address that was flooded (L23). Time related topics were also prominent (T1-6) that included day and year of Sandy's landfall (October 29, 2012) in addition to other specific time units.

Media played a salient role during Hurricane Sandy. Numerous patterns were part of the media coverage, both traditional news media (NM1-16) and easily accessible social media (NM17-25). Traditional media sources included CNN, LA Times, Sky News, NYT, ABC News, Daily Mail, CBS, USA Today, Washington Post, Huffington Post, BBC, Forbes, Reuters, and Wall Street Journal. On the other hand, Facebook, Twitter, and Instagram were among the popular social media topics as people talked about several social media activities such as 'tweet', 'status', 'page', 'share', 'message', 'follow' and so on. Some topics represented solely different broadcasting steps such as live updates, live coverage, press briefings, and others (NM26-30). Political leaders such as President Obama (PC1-3), Mitt Romney and Bill Clinton (PC4-7) appeared in several discussion topics. Some interactions were on the 2012 US Election and its relevance to Hurricane Sandy (PC8-12). Celebrity activities during that period were also prominent in a number of occasions. For example, Karl Rove, Jim Cantore, Lindsay Lohan, and Dina Lohan appeared in a few conversations (PC13-16). Some other topics were uncategorized (O1-33).

## DISCUSSION

Conversations of most frequent and relevant users (~763 K tweets from top 4029 highly active users; $AF \geq 100$) indicate evolution of numerous topics at different phases of the storm such as warning, response, and recovery. People were also concerned about phase independent topics specific to location, time, media coverage, political leaders, and celebrity activities. We observed that the top 50 most frequent words contribute up to ~30% of all words in those tweets, however, each word can be part of any specific topic or multiple topics. Warning phase topics were related to storm prediction, storm watch and advisory, storm preparedness, user concerns about the storm, previous hurricane experience and causes responsible for storm formation such as climate change and global warming. Communications during the response phase may include basic needs of people such as food, water, gas, power and so on. People also express concern on major infrastructures and different transportation facilities such as subway, metro, train, bus, airlines, and others. Many interactions include how local officials responded to the storm threat and order mandatory evacuations. First responders such as FEMA, hospitals, pets/animals, fire, flooding, event cancellation, rescue operations, work or school closure were among the topics. People primarily discussed about several disaster relief and fundraising efforts in the recovery phase. In addition, possible aftermath and lessons learned from such major disaster also emerged during this phase. While some communication patterns are location (impacted states/areas) and time specific, some were specific to media coverage including both traditional news media and social media. Political leaders, 2012 US election and celebrity activities were among other topics of discussion.

The pattern recognition approach of this study can be used in identifying real time user needs in future crises. For example, if users express their concern about locating nearby shelters in social media, this will emerge out if the model is applied. Introducing hashtags are common these days in social media. The keywords that constitute a particular pattern of user concern, as presented in this study, would help emergency managers to use such keywords as hashtags at different phases (warning, response and recovery) of major hurricanes and help vulnerable population receive more crisis information via social media. Future studies should capture the complex contagion of hurricane Sandy in more details. For example, how user concerns spread through the network of twitter users (followee-follower versus direct user mentions) and the rate of information propagation, content relevance, risk profiling. Subgraphs of geo-tagged users (~2% in the raw data) may help to understand the mobility pattern and spatial correlation with their network activity.

## ACKNOWLEDGEMENT

The authors are grateful to National Science Foundation for the grant CMMI-1131503 and CMMI- 1520338 to support the research presented in this paper. However, the authors are solely responsible for the findings presented in this study.

Sadri, Hasan, Ukkusuri, Cebrian																																																																																																														18